\documentclass[showpacs,preprintnumbers,english,twocolumn,amsmath,amssymb]{revtex4-1}
\usepackage[T1]{fontenc}
\usepackage[latin1]{inputenc}
\usepackage{graphicx}
\usepackage{amssymb}
\usepackage{verbatim}
\usepackage{epic,eepic}
\usepackage{babel}
\usepackage{lmodern}
\usepackage{color,soul}
\usepackage{epsfig}
\usepackage{txfonts}
\usepackage{subfigure}
\usepackage{float}
\usepackage{wasysym}
\usepackage{url}
\usepackage{hyperref}
\usepackage{rotating}

\newcommand{\mn}{\ensuremath{\overline{\nu}}\,}
\newcommand{\emean}{\ensuremath{\overline{E}}}
\newcommand{\pu}{\ensuremath{^{239}}Pu\,}
\newcommand{\jeff}{JEFF3.3 }
\newcommand{\ENDF}{ENDF/B-VIII.0 }
\date{}
\begin{document}

\title{Prompt Fission Neutron Spectra in the $^{239}$Pu($n,f$) Reaction}

\author{P.~Marini$^{1}$}\email{paola.marini@cea.fr}
\author{J.~Taieb$^{1}$}
\author{B.~Laurent$^{1}$}
\author{G.~Belier$^{1}$}
\author{A.~Chatillon$^{1}$}
\author{D.~Etasse$^{2}$}
\author{P.~Morfouace$^{1}$}
\author{M.~Devlin$^{3}$}
\author{J.~A.~Gomez$^{3}$}
\author{R.~C.~Haight$^{3}$}
\author{K.~J.~Kelly$^{3}$}
\author{J.~M.~O'Donnell$^{3}$}
\author{K.~T.~Schmitt$^{4}$} 

\affiliation{
$^{1}$CEA, DAM, DIF, F-91297 Arpajon, France\\
$^{2}$LPC Caen, ENSICAEN, Universit\'e de Caen, CNRS/IN2P3, Caen, France \\
$^{3}$P-27, Los Alamos National Laboratory, Los Alamos, NM-87545, USA\\
$^{4}$ISR-1, Los Alamos National Laboratory, Los Alamos, NM-87545, USA
}

\date{\today}

\begin{abstract}
Prompt fission neutron spectra from \pu($n,f$) were measured with respect to $^{252}$Cf spontaneous fission for incident neutron energies from $0.7$
to $700\,$MeV at the Weapons Neutron Research facility (WNR) of the Los Alamos Neutron Science Center. 
A newly designed high-efficiency fission chamber was coupled to the highly segmented Chi-Nu neutron liquid scintillator array to detect neutrons emitted in fission events.
The double time-of-flight technique was used to deduce the  incident-neutron energies from
the spallation target and the outgoing-neutron energies from the fission chamber. Prompt fission neutron spectra (PFNS) were measured with respect to $^{252}$Cf spontaneous fission down to $200\,$keV and up to about $12\,$MeV for all the incident neutron energies with typical total uncertainties well below $2\%$ up to about $7\,$MeV outgoing-neutron energy. 
The general trend of PFNS is well reproduced by \jeff and \ENDF evaluations, although a better agreement is found with \jeff. Discrepancies were  observed for the low-energy part of the spectra, especially around the opening of the $2^{nd}$-, $3^{rd}$- and $4^{th}$-chance fission. 
Neutron 
average kinetic energies as a function of incident-neutron energy are obtained experimentally with  reported total uncertainties below $0.5\%$. 
The measured values 
agree with the most recent data. 
The trend is fairly well reproduced by the \jeff evaluation, although it 
fails to reproduce the experimental values within their uncertainties.\\
\end{abstract}


\maketitle

\section{Introduction}
Observables related to prompt fission neutrons
represent a key parameter for nuclear energy applications and technology, and provide valuable information on the fission process. 
As far as the applications are concerned, it has been recently shown that accurate predictions of nuclear criticality using
neutron transport codes crucially depend on the underlying
nuclear data, especially the prompt fission neutron spectrum \cite{peneliau2014}.
While the accuracy of evaluated  neutron multiplicities, \mn, 
has improved in recent years for some nuclei, the 
experimental database of prompt fission neutron spectra  consists of a limited number of datasets, far less precise than those on the other key fission observables  and often discrepant \cite{capote2008, neudecker2016}.

For the $^{239}$Pu($n,f$) prompt fission neutron spectrum (PFNS) the situation is similar: the few experimental data sets  that are available and adequate for an evaluation \cite{nefedov, boytsov, starostov, 
starostov89, lajtai, staples1995,  knitter75,lestone2014,chatillon, noda2011} are often not  in agreement within quoted uncertainties \cite{neudecker2016,capote2016,neudecker2018}. 
Moreover, there are large gaps in the experimental data for important parts of the outgoing-neutron spectra, i.e., below $1\,$MeV and above $8\,$MeV.  
These two regions of the PFNS are particularly challenging because of technical limitations (detection efficiency at low outgoing-neutron energy and statistics at high energy). The high $\alpha$ activity of the $^{239}$Pu isotope makes the measurement even more challenging to carry out.

Due to these deficiencies in experimental data, evaluated data files partially rely on models to predict the spectra. 
However, despite the considerable progresses in recent years, microscopic  models are not yet able to precisely predict the fission observables, and in particular PFN spectra, because of the complex physical processes associated with the prompt emission  of neutrons from fission of actinides. Phenomenological models, mainly based on the Madland-Nix approach \cite{madland_LosAlamosModel1}, are then used to evaluate PFN spectra in the libraries. These models rely on  parameters, tuned as best fits to the measured PFN spectra. 
However, it has been suggested that significant systematic errors could be present in the evaluated libraries, leading to a too hard PFNS \cite{chadwick2011}. In particular, according to 
 Ref. \cite{maslov2011}, the PFN spectra of major actinides 
 should have more neutrons below $1\,$MeV 
 and fewer neutrons above about $6\,$MeV outgoing-neutron energies.
 Recently developed Monte Carlo Hauser-Feshbach models \cite{talou2011, freya, fifrelin} point to the same direction.
For the  fast-neutron induced $^{239}$Pu($n,f$) reaction, recent data from Lestone and Chatillon \cite{lestone2014,chatillon, noda2011} support current PFNS evaluations for outgoing energies between $1.5$ and $10\,$MeV, and $0.5$ and $8\,$MeV, respectively.
But these  data  do not 
rule out the ideas proposed in Ref. \cite{maslov2011} that there should be more prompt fission neutrons below 1 MeV and fewer above  $10\,$MeV.

The open question on the high-energy tail of the spectrum has important implications also on the understanding of the fission process. Indeed, this part of the spectrum  is very sensitive 
to the total excitation energy sharing at scission: the spectrum gets harder
at higher excitation energy. 
While in the original Madland-Nix model \cite{madland_LosAlamosModel1, madland_LosAlamosModel2} a thermal equilibrium between the two nascent fragments at scission is assumed, 
a recent model \cite{karlheinz2010,karlheinz2011,karlheinz2011_2} describes the fission
process with a constant temperature level density, where each nascent
fragment would be characterised by a temperature only dependent on the mass number of the fragment.
This model 
naturally leads to an unexpected partitioning of the excitation energy when increasing the incoming neutron energy \cite{karlheinz2010}, which would affect the high-energy tail of the spectrum.

Our group has been involved since the 2000's in PFNS measurements \cite{ethvignot,noda2011,chatillon,taiebProc,keegan_Pu}, in the framework of a collaboration  agreement between the US Department of Energy - National Nuclear Security Administration (DOE/NNSA) and the French  Commissariat \`a l'\'energie atomique et aux \'energies alternatives - Direction des applications militaires (CEA-DAM).
First experiments on PFN spectra of $^{239}$Pu($n,f$) were carried out in 2007 and 2008 \cite{noda2011,chatillon}. Since then special efforts in the development of a new experimental area \cite{devlin2018}, new detectors, and
data collection procedures were made with the aim of improving the measured
PFNS accuracy and precision. 
In particular, a new fission chamber was developed, with a fission fragment detection efficiency better than $95\%$, despite the high $\alpha$-activity of the sample. Such a feature is crucial to avoid any bias of the data associated to a particular range in angle or total kinetic energy.
 In this work, we present the results of the  $^{239}$Pu($n,f$) PFNS measurements performed in 2017 at the Los Alamos National Laboratory (LANL). 
The goals of this experiment are: (i) to significantly improve
the experimental accuracy of the measured PFNS, 
in particular in the low- and high-energy tails of the spectrum, in the $1\,$ to $700\,$MeV incident neutron energy range; (ii) to produce PFNS data as a function of incident-neutron energy, for which few experimental data exist to date, (iii) to provide an independent measurement of emitted-neutron 
mean kinetic energy as a function of incident-neutron energies with high accuracy.

The paper is structured as follows: in Secs.\ref{sec:experimental setup} and \ref{sec: data analysis} the experimental setup and the data analysis are presented, respectively. Section \ref{sec:recults} presents and discusses the obtained results, while Sec. \ref{sec:conclusions} concludes.


\section{Experimental setup}\label{sec:experimental setup}

The experiment was carried out at the WNR facility \cite{wnr1,wnr} at  Los Alamos National Laboratory.
A white pulsed neutron beam, spanning an energy range from below $1\,$MeV to several hundreds of MeV, is produced by $800\,$MeV proton-induced spallation reactions  on a tungsten target. 
The beam is delivered in $625\,\mu$s-long macropulses with an average frequency varying between $100$ and $120\,$Hz. Each macropulse contains  micropulses separated by $1.8\,\mu$s. 
A $5\%$-borated, $1.27\,$cm-thick polyethylene absorber 
is placed into the beam line 
to harden the spectra in order to reduce the so-called wrap-around effect, i.e., to minimize the number of slow neutrons ($\lesssim680\,$keV) reaching the fission target at the same time or after the fastest neutrons of the following pulse.
The neutron beam is collimated along a flight path of about $20\,$m before impinging onto the \pu fission target. A beam pipe right upstream from the experimental hall was placed under vacuum to limit neutron scattering on air.

\begin{figure}[t!]
\centering
\includegraphics[width=0.9\columnwidth,clip]{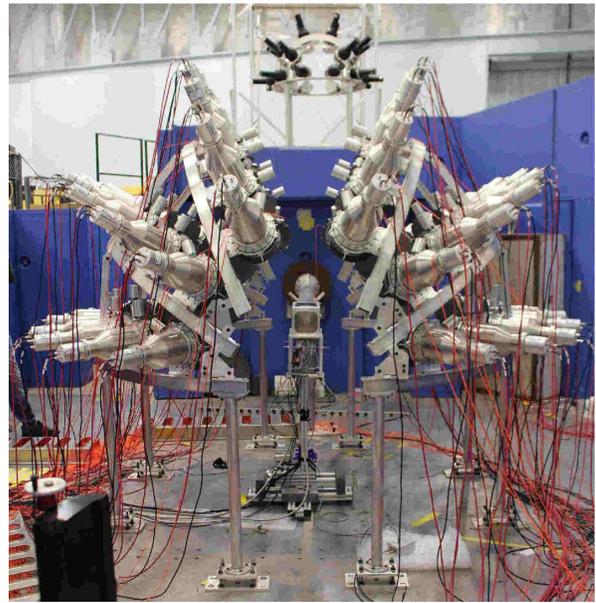}
\caption{Photo of the experimental setup. The photo is looking upstream towards the spallation target. The fission chamber is in the center of the photo and
is viewed by $54$ liquid scintillators. The floor is $1.25\,$cm-thick aluminum over a
get-lost pit $2$ meters deep.}
\label{fig:setup}       
\end{figure}
\begin{figure}[h!]
\centering
\includegraphics[width=0.9\columnwidth,clip]{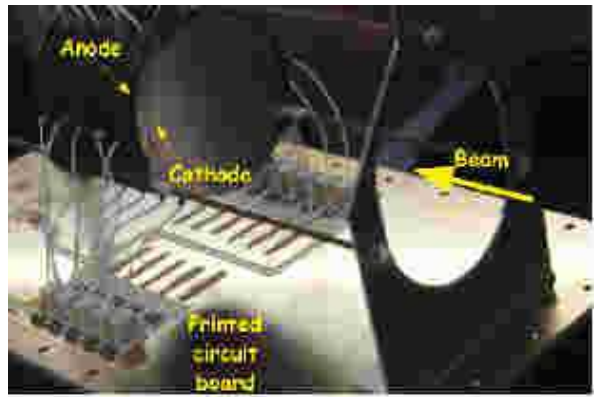}
\caption{Photo of the interior of the fission chamber, showing the stacking of anodes and cathodes. The cathodes contain the \pu deposits, the anodes are connected to the lower panel. }
\label{fig:fc}       
\end{figure}
Data were taken for 20 days  of effective time with a fission rate of $150\,$s$^{-1}$.

\begin{figure}[h!]
\centering
\includegraphics[width=0.6\columnwidth,clip]{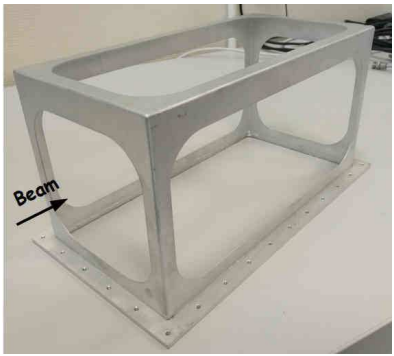}
\caption{Photo of the low-mass $1\,$mm-thick aluminum outer structure of the fission chamber. $50$ and $100\,\mu$m-thick tantalum foils glued on the sides ensure the airtightness of the chamber \cite{taieb}.}
\label{fig:fc box}       
\end{figure}

The experimental setup (Fig. \ref{fig:setup}) couples the $54$ Chi-Nu liquid scintillator  array \cite{chinu} to a newly developed fission chamber and a digital data acquisition system. The detectors are placed over a get-lost pit to minimize the amount of scattered neutrons on surrounding materials (see Sec. \ref{sec:background corrections}).

\subsection{Fission chamber}
\label{sec:fission chamber}
The multi-plate fission chamber (Fig. \ref{fig:fc}) was designed to contain  $47\,$mg of \pu  arranged in $22$ deposits and $11$ readout channels. The fissile material was deposited  on every cathode, located  $2.5\,$mm away from its corresponding anode. The diameter of each deposit was $33\,$mm and the isotopic purity of the material $99.90\%$.
The chamber, filled with CF$_{4}$ gas to ensure a fast charge collection,   operates $100\,$mbar above the local atmospheric pressure. This allows  for a reduction in the amount of structural material, as shown in Fig. \ref{fig:fc box}, thereby minimizing the neutron scattering both for the incoming neutron beam and for the emitted fission neutrons (see Sec. \ref{sec:background corrections}).
 A more detailed description of the detector can be found in \cite{taieb}.

\begin{figure}[t!]
\centering
\includegraphics[width=0.95\columnwidth,clip]{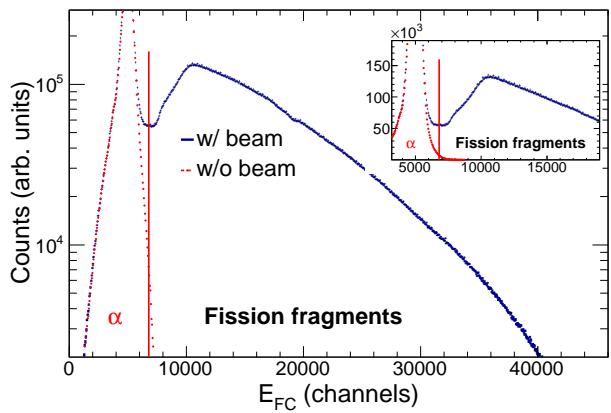}
\caption{(Color online) Pulse height spectra of \pu fission chamber obtained with (blue) and without (red) beam.  The
vertical red line indicates the cut used in the data analysis to separate $\alpha$-particles from fission fragments. 
(see Sec. \ref{sec: alpha-FF threshold}).}
\label{fig:FC}       
\end{figure}

Front-end electronic boards containing  pre-amplifiers and shapers were developed to match the detector characteristics.
Thanks to the detector and electronics design,  
an improved $\alpha$-fission fragment discrimination could be obtained, with a fission detection efficiency better than $95\%$, despite approximately $10\,$MBq $\alpha$ activity per channel \cite{budtz}. Two pulse height spectra (shown in Fig. \ref{fig:FC}) measured during the experiment with (in blue) and without (in red) beam illustrate the discrimination obtained.
A time resolution better than $0.8\,$ns full width at half maximum (FWHM) could be achieved. This leads to an
emitted-neutron
 time-of-flight  resolution of  about $1.5\,$ns (FWHM). 
 The latter arises 
 from summed contribution of the fission chamber  and  the neutron detector time resolutions. 

An identical chamber, containing a $^{252}$Cf deposit of the same size as the \pu ones, was  used for neutron detection efficiency measurements (see Sec. \ref{sec:neutron det eff}). \\

For sake of completeness, we mention here that, during the experiment two channels of the Pu fission chamber were not working. Those are the third and the seventh channels. This does not introduce any significant bias in the experimental results, since the interaction probabilities of $1$ and $10\,$MeV fission neutrons with the material supporting each deposit are $0.3\%$ and $0.2\%$, respectively. In other words,  the stack of \pu deposits is equivalent to a single deposit, with the advantage of a thin deposit, i.e., allowing the recoil of fission fragments out of the target.

\subsection{Neutron detection array}
\label{sec:Chi-Nu}
The Chi-Nu neutron detector array \cite{chinu} was used to measure neutrons in coincidence with a fission chamber signal.
The Chi-Nu high-energy array consists of 54 EJ-309 liquid scintillator cells, read by photomultipliers, and arranged on a hemisphere with inner radius of about $1\,$m around the fission chamber, as shown in Fig. \ref{fig:setup}.
The diameter and the thickness of each cell are 
$17.78\,$ and $5.08\,$cm, respectively. 
 The whole array covers close to $10\%$ of the solid angle. 
 The detectors are placed at nine different angles, with $6$ detectors per angle, $\theta_{lab}$, from $30^{\circ}$ to $150^{\circ}$, with respect to the beam axis. The angular distribution of the emitted neutrons can therefore be studied. The use of the Chi-Nu array constitutes a significant improvement with respect to previous measurements \cite{noda2011,chatillon} performed with the FIGARO array, due to the higher number of scintillation detectors
and the significant reduction of scattering material near the fission chamber and the detector array.

The EJ-309 liquid scintillator is sensitive to both neutrons and $\gamma$-rays. Particle identification was performed 
via Pulse Shape Discrimination (PSD) technique, which is based on the charge integration of the slow and  fast components of the pulse. The neutron-$\gamma$ discrimination was optimized adjusting the duration of the time integration windows.

\begin{figure}[t!]
\centering
\includegraphics[width=0.95\columnwidth,clip]{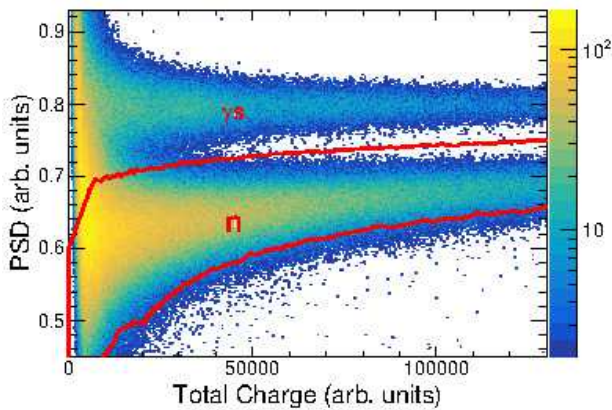}
\caption{(Color online) Correlation between the pulse shape discrimination (PSD) and the total charge deposited in 
an EJ309 scintillator by $\gamma$ rays and neutrons. Signals falling into the area outlined by
the red contour are associated to the detection of neutrons. 
(see Sec. \ref{sec: ng discri}).} 
\label{fig:PSA}      
\end{figure}

A $2\,$mm-thick Pb shield was placed in front of each cell to reduce the amount of low energy ($\lesssim300\,$keV) $\gamma$-rays, impinging on the cell. 
A typical pulse-shape discrimination plot is shown in Fig. \ref{fig:PSA}, where the slow to fast pulse-component ratio (PSD) is plotted as a function of the total pulse charge. The use of the Pb shield allowed us to obtain a clear neutron-$\gamma$ discrimination down to $30-40\,$keV electron equivalent energy  \cite{pino2014}.\\

\subsection{Data acquisition system}
\label{sec:daq}
The digital Fast Acquisition System for nuclEar Research (FASTER)  \cite{faster} was used during the experiment. Signals were digitized by a $500\,$MHz, 12-bits, low noise Analog to Digital Converter (ADC) and processed by real time numerical modules implemented into Field Programmable Gate Arrays (FPGA). The signal-to-noise ratio, as well as the zero time crossing determination on the Constant Fraction discriminator (CFD) signal, are optimized by limiting the analog bandwidth with the use of an input passive low pass filter ($100\,$MHz).
This allows to obtain a time resolution as low as $7.8\,$ps.
The use of this acquisition system allowed the near complete avoidance of numerical dead-time. 
The acquisition was triggered and a coincidence window of $1.82\,\mu$s opened when a signal was present in the fission chamber. Signals from the scintillators falling in the coincidence window were recorded. 

\section{Data analysis}\label{sec: data analysis}
The experiment is based on the double time-of-flight technique. For each triggered event, the time of flight of the incoming neutron from the spallation target to the fission chamber and the time of flight of the emitted neutron from the fission chamber to the scintillator were measured. 
From the measured raw time of flight, the absolute times of flight for incoming and emitted neutrons were determined from the position of the $\gamma$-flash generated by the interaction of the proton beam on the tungsten target and from prompt-fission $\gamma$-rays, respectively.
Distances between the fission target and each scintillator cell
were carefully measured with an uncertainty better than $0.2\%$ with a high precision laser range meter. 
Given the importance of knowing the distance precisely, a careful consistency check of the  measured   values  was performed at the very beginning of the data analysis.

To measure the flight path length from the spallation source to each deposit in the fission chamber, a $0.5\,$inch-thick carbon
absorber was inserted in the beamline upstream the experimental hall. The positions of the $2.08\,$ and $6.3\,$MeV resonances in the total cross section gave the distances between each deposit of the fission chamber and the spallation target, with an absolute uncertainty of $4\,$mm over the $21.5\,$m flight path.

The combination of the time-of-flight and distance measurements gives us access to the incident and emitted neutron kinetic energies.
The obtained relative uncertainty on the incident-neutrons energy is $0.08\%$. The uncertainty on the outgoing-neutron energy is dominated by the time-of-flight resolution ($0.6\,$ns $1\,\sigma$).
\begin{figure}[t!]
\centering
\includegraphics[width=0.95\columnwidth,clip]{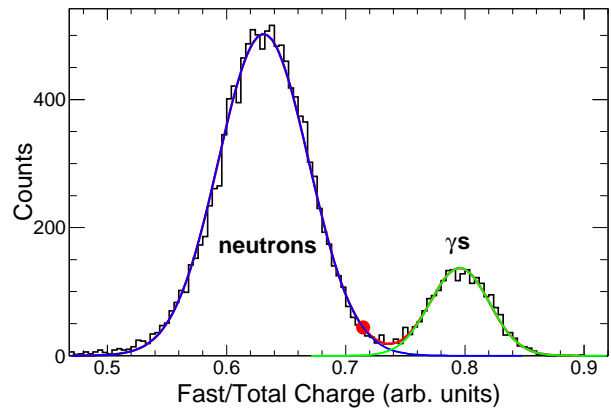} 
\caption{(Color online) Example of Gaussian fit of the fast to total pulse components ratio for a given charge  interval measured in one cell.  The dotted blue curve is the fit of the neutron component, while the green line corresponds to the $\gamma$-ray component. The red dot indicates the position of the used threshold to discriminate neutrons and $\gamma$-rays.
Selections were applied neither on incident-neutron energy nor on emitted-neutron energy, but were used to evaluate the remaining contribution of the $\gamma$-rays
 (see text).} 
\label{fig:PSA gaus fit} 
\end{figure}

\subsection{Events selection}\label{sec:Events selection}
\subsubsection{$\alpha$-fission fragments discrimination}\label{sec: alpha-FF threshold}
A threshold was set to discriminate fission from $\alpha$-decay and nuclear reaction events
in the fission chamber pulse-height spectrum (see Fig. \ref{fig:FC}). 
The threshold position is the best compromise between minimizing the contribution of $\alpha$-decay and reaction events and preserving the full fission-fragment total kinetic energy and angular distributions.
\subsubsection{$\gamma$-neutron discrimination}\label{sec: ng discri}
Special efforts were made to improve the $\gamma$-neutron discrimination and $\gamma$-ray rejection.
A first selection based on the detected time-of-flight was performed to remove prompt $\gamma$-rays.
An additional selection was placed, for each neutron detector, on the PSD vs total  charge correlation, shown in Fig. \ref{fig:PSA}, to remove the remaining $\gamma$-rays.
The cut was determined  according to the procedure described in Ref. \cite{polack}: for each total  charge interval, the fast to total pulse-component ratio was fitted with two Gaussians, as shown in Fig. \ref{fig:PSA gaus fit}. This allowed us to define the cut with the same criterion for all the detectors and to estimate the remaining contribution of $\gamma$-rays, which is lower than $0.1\%$.\\
\begin{figure}[t!]
\centering
\includegraphics[width=0.95\columnwidth,clip]{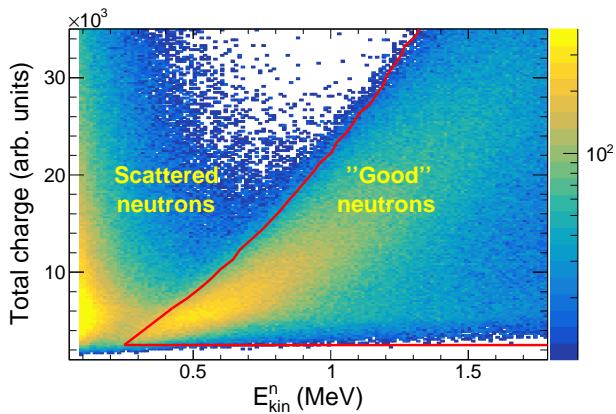}
\caption{(Color online) Correlation between  scintillator light output and outgoing-neutron kinetic energy, deduced from the time-of-flight measurement, of  detected neutrons. The red contour selects ``good neutrons'' (see text).}
\label{fig:PSA Light vs Ekin} 
\end{figure}
To remove a large fraction of scattered neutrons impinging on the detectors, a light output vs. outgoing-neutron kinetic energy correlation, as shown in Fig. \ref{fig:PSA Light vs Ekin}, was made. Since a  correlation exists between the  kinetic energy of the neutron impinging on the scintillator cell and the scintillator light output, only events satisfying this correlation were selected (red contour in Fig. \ref{fig:PSA Light vs Ekin}).

With these constraints, neutrons down to $200\,$keV could be identified. As these selections are applied both to the \pu and $^{252}$Cf PFN spectra analyses as discussed below, possible systematic uncertainties associated to the selections cancel out when correcting the measured \pu spectra for the neutron detector efficiency (see Sec. \ref{sec:neutron det eff}).

\subsection{Neutron detector efficiency}\label{sec:neutron det eff}
\begin{figure}[t!]
\centering
\includegraphics[width=0.95\columnwidth,clip]{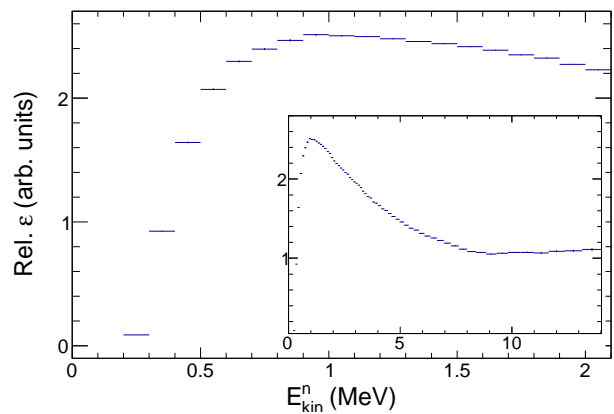}
\caption{Measured mean relative efficiency as a function of the emitted neutron energy. The horizontal bars indicate the outgoing-neutron kinetic-energy bin-width. In the insert is shown the whole curve.} 
\label{fig:efficiency} 
\end{figure}

Neutron detector efficiencies were obtained by measuring the PFNS of the $^{252}$Cf spontaneous fission reaction in the same experimental conditions. 
\begin{figure}[ht]
\centering
\includegraphics[width=0.95\columnwidth,clip]{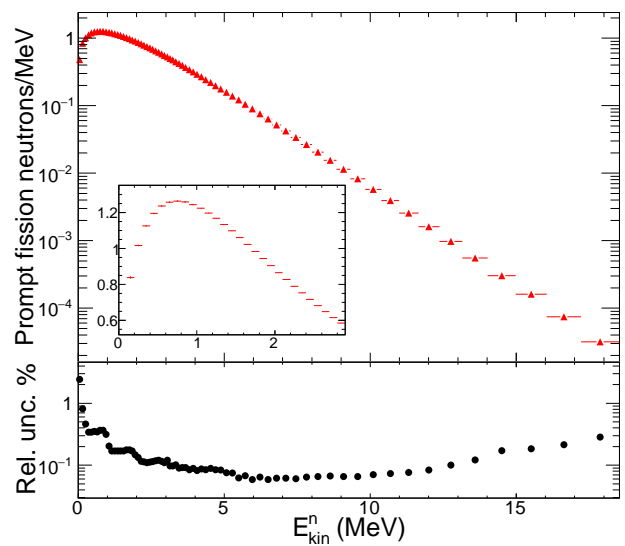}
\caption{Evaluated $^{252}$Cf PFNS \cite{carlson2009}, normalized to $3.7637\,$ neutrons \cite{carlson2018}, used for neutron detection efficiency determination (see text). Horizontal bars indicate the width of the outgoing-neutron kinetic energy bins. In the insert a zoom of the low energy part of the spectrum is presented.
 The relative uncertainties are plotted in the lower panel. }  
\label{fig:CFspectrum} 
\end{figure}
For this purpose,  beam-off runs   were performed with a dedicated fission chamber containing a 
 $^{252}$Cf deposit placed on the central cathode of the chamber. All the structural materials constituting the \pu fission chamber were present in the chamber used for the $^{252}$Cf measurement. Fission signals triggered the data acquisition.
The presence of a single deposit in the Cf chamber, as contrasted to the 11 deposits in the Pu chamber, does not introduce a significant distortion of the measured spectra due to the very low interaction probabilities of fission neutrons with the material supporting each deposit. Moreover the exact distance of each Pu deposit to each neutron detectors was accounted for.
The efficiencies were  determined from $^{252}$Cf spontaneous fission PFNS with respect to the Mannhart
 evaluated PNFS standard \cite{carlson2009}, shown in Fig. \ref{fig:CFspectrum},  normalized to the evaluated standard $\overline{\nu_{tot}}$ of ($3.7637\pm0.42\%$) for $^{252}$Cf(sf) 
  \cite{carlson2018}.
An example of the typical relative neutron detection efficiency is shown in Fig. \ref{fig:efficiency}. We stress that a threshold as low as $200\,$keV neutron energy was  obtained with the present setup. The counting statistics for the $^{252}$Cf(sf) was of $2.7\cdot10^{8}$ fission events, comparable to the one collected for the \pu(n,f).
\subsection{Scattering corrections}
\label{sec:background corrections}
The presence of structural and surrounding materials, and to a lesser extent air, causes the scattering of neutrons distorting the measured PFNS. 
The whole setup was  placed above a $2\,$m deep ``get-lost'' pit, the beam pipe  upstream from the experimental hall was placed under vacuum, and special efforts were dedicated to the development of the fission chamber with a reduced amount of scattering material (see Sec. \ref{sec:fission chamber} and Fig. \ref{fig:fc box}). 
Nevertheless, the remaining scattered neutron contributions have to be corrected for.
\begin{figure}[b]
\centering
\includegraphics[width=0.95\columnwidth]{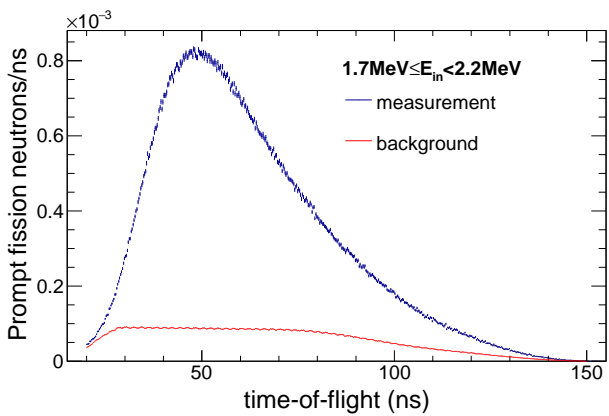} 
\caption{(Color online) time of flight of emitted neutrons (blue) and associated background (red) measured for an incident neutron energy between  $1.7$ and $2.2\,$MeV and normalised to the number of  events.} 
\label{fig:tof n inc} 
\end{figure}
Scattering of prompt fission neutrons leads to an erroneous determination of their flight path, and thus of their kinetic energy. This contribution is accounted for, at a first order,  when correcting the spectra for the detector efficiency. Indeed the efficiency measurement was performed in the same experimental conditions as the real measurement. Moreover, under the reasonable assumption that the $^{252}$Cf and \pu have similar PFN spectra, the distortions due to the remaining scattering events are compensated for in the efficiency calibration process. This is further discussed in Sec. \ref{sec:simu}.

The most significant contribution to spectrum distortion and background comes from neutrons scattered from the incident neutron beam and introduces a significant background in the PFNS.
This background is not correlated with fission events,  thus creating random coincidences in neutron detectors, and its contribution varies with time after the proton burst. 
A 
 pulser trigger, acting as a fake-fission generator, 
 was used to start the acquisition and record neutron-detector random coincidences during the whole experiment. 
As the beam was on, the various incident-neutron times-of-flight
were randomly sampled and 
the background monitored as a function of the beam energy. 
The pulser rate was much higher than the rate of real fissions ($\simeq 150\,$fissions$\cdot\,$s$^{-1}$), so that the statistical uncertainty associated to the background contribution was $4$ times smaller than the statistical uncertainty associated to real fissions. An example of measured fission neutrons and random coincidence times-of-flight is shown in Fig. \ref{fig:tof n inc} in blue and red, respectively. It can be noted that for times-of-flight under about $20\,$ and above $150\,$ns (corresponding to energies of about $13\,$ and $0.2\,$MeV, respectively) the difference between the measurement and the background vanishes, validating the correct background measurement and subtraction.
Random coincidences  were then subtracted offline.
\subsection{Extraction  of the PFNS}\label{sec: analysis PFNS}
\begin{figure}[b]
\centering
\includegraphics[width=0.95\columnwidth,clip]{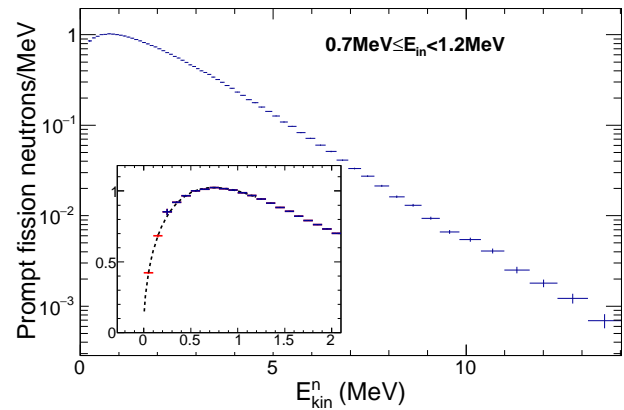}
\caption{(Color online) PFNS for an incident neutron energy ranging from $0.7$ to $1.2\,$MeV. When not visible uncertainties are smaller than symbols. In the insert is a zoom of the low energy part of the experimental spectrum (blue). The dashed line is the Watt-type fitted function and the red symbols are the extrapolated values below $200\,$keV.}
\label{fig:pfns with fit watt} 
\end{figure}
Thanks to the high statistics  accumulated during the experiment and the low detection thresholds, PFNS from $200\,$keV to about $15\,$MeV could be extracted, the associated PFN observables  studied, and their evolution investigated as a function of the incident neutron energy. 
Data were  sorted in eighty-six incident neutron energy bins with energies ranging from $0.7$ to $750\,$MeV. 
 The bin widths were chosen based on the available statistics and on the incident-neutrons time-of-flight resolution. The bin width was $0.5\,$MeV at low incident neutron energies and it increased up to $20\,$MeV above $250\,$MeV. The used bin-widths are reported in Table \ref{tab:bin width}.
  \begin{table}
\begin{center}
\begin{small}
\begin{tabular}{|c|c||c|c|}
\hline
E$_{in}$ range & E$_{in}$ bin width & E$_{in}$ range & E$_{in}$ bin width\\
(MeV) &  (MeV) & (MeV) &  (MeV)\\
\hline
 $0.7-6.2$      & $0.5$ & $26.4-32.4$  & $3.0$\\
 $6.2-9.4$    & $0.8$ & $32.4-82.4$  & $5.0$\\
 $9.4-20.4$   & $1.0$ & $82.4-252.4$  & $10.0$\\
 $20.4-26.4$  & $1.5$ & $>252.4$  & $20.0$\\
\hline
\end{tabular}
\end{small}
 \end{center}
\caption{Incident-energy bin widths used in the analysis.}
\label{tab:bin width}
\end{table}
   
For each detector, time-of-flight spectra were measured for
each  incident neutron energy bin. The time spectra, corrected for random coincidences, were then
converted into energy spectra and corrected for the detection efficiency normalized to the $^{252}$Cf(sf) 
 standard. The final spectrum for a given incoming-neutron energy range was finally obtained by combining all the detector spectra. 
In the following,  results are presented with the absolute statistical and systematics
error bars, propagated through the data analysis. 
The systematic uncertainty is the uncertainty on the evaluated $^{252}$Cf spectrum.
The horizontal bars indicate the width of the emitted neutron energy bin and do not represent an error bar. The chosen variable bin width for the energy spectra is a compromise between a precise shape of the spectrum, enough statistics and the outgoing-neutron time-of-flight resolution.

An example of the measured prompt fission neutron spectra is shown in Fig. \ref{fig:pfns with fit watt}. The spectrum was obtained for an incident neutron energy ranging from $0.7$ to $1.2\,$MeV, with a mean energy of $0.97\,$MeV. 
The performances of the setup and the associated electronics chain, and the high collected statistics, allowed PFN
 spectra to be experimentally measured 
from $0.25\,$MeV to $10$-to-$12\,$MeV, depending on the incident neutron energy, with total uncertainties from $0.2\%$ to $10\%$ around $0.8\,$MeV and $12\,$MeV, respectively.

\subsubsection{Extrapolation of PFNS at low kinetic energy}\label{sec:extrapolation of pfns at low ek}
Despite the rather low outgoing-neutron energy threshold, it should be kept in mind that quantities such as the mean kinetic energy 
of prompt fission neutrons, \emean, 
are sensitive to the low-energy part of the spectrum. 
 Owing to simple theoretical descriptions of PFNS based on either Watt \cite{watt1952} or Maxwellian \cite{bloch1943}  spectra, the measured spectra were  extrapolated at energies from $0.2\,$MeV downwards by fitting the low energy part of the spectra with a Watt-type 
   and a Maxwell-type distribution  \cite{terrell1959}. The Watt-based extrapolation was limited to incident energies below the opening of the second-chance fission, around $6\,$MeV.
 The fitting ranges were from $0.2$ to $1.2\,$MeV for the two functions. The uncertainties on the extrapolated points at $0.05$ and $0.15\,$MeV were given by the uncertainties and covariances on the fit parameters.
 Consistent results were obtained below $6\,$MeV incident-neutron energy when using Watt-type and Maxwell-type distributions. This validates the use of a Maxwell-type distribution at energies above $6\,$MeV.
As a test case, the same extrapolation method was applied to identically truncated PFNS generated by the General Description of Fission Observables (GEF) code \cite{GEF}. The procedure was applied to GEF PFNS from $1$ to $20\,$MeV incident neutron energies. From the comparison of the extrapolated and the full distributions, we observe differences smaller that $0.1\%$ on the PFN spectra, as well as on the 
PFN mean kinetic energy, \emean, thus validating the extrapolation method.
An example of the result is shown in the insert in Fig. \ref{fig:pfns with fit watt}, where blue points are experimentally measured and red points are extrapolated.
In the following these extrapolated spectra will be used, unless differently specified.

\subsection{Study of possible experimental bias}\label{sec:simu}
\begin{figure}[t!]
\centering
\includegraphics[width=0.95\columnwidth,clip]{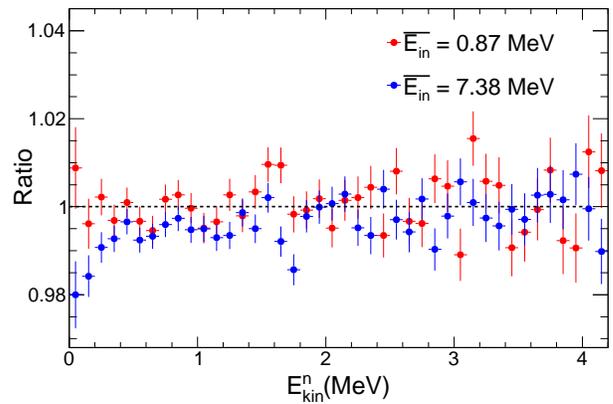}
\caption{(Color online) Ratios between the $^{239}$Pu(n,f) PFN spectrum determined as in the data analysis and the PFN spectrum used as input in the simulation for $0.87$ (red) and $7.38\,$MeV (blue) incident neutron energies. .}
\label{fig:simu pierre} 
\end{figure}
As mentionned neutron distorsion spectra corrections are obtained through the measurement of  neutron detection efficiencies with a $^{252}$Cf fission chamber (see Sec. \ref{sec:neutron det eff}). 
Under the reasonable assumption that the $^{252}$Cf PFNS is sufficiently close to the $^{239}$Pu, this is justified since the $^{252}$Cf and the $^{239}$Pu fission chambers are exactly the same. In order to validate 
%
%
it the fission chamber as well as neutron detectors were simulated with the GEANT4 toolkit \cite{geant4} using the NPTool package \cite{matta}. The custom neutron interaction model, MENATE$\_$R, was used, where the inelastic neutron-carbon and neutron-hydrogen reactions are modeled as discrete reaction channels based on experimental information  \cite{kohley_menate}.
 The input $^{252}$Cf PFNS was assumed to be the evaluated Mannhart's spectrum \cite{carlson2009}, while experimentally measured PFN spectra for incident mean energies of $1$, $7.4$ and $14\,$MeV were used as input for $^{239}$Pu(n,f) PFN spectra. Fifty million events were generated and the simulated $^{252}$Cf(s.f.) and $^{239}$Pu(n,f) PFN spectra extracted. 
  The detector efficiencies and the $^{239}$Pu spectra were determined following the same procedure as for the experimental data. 
  
  The influence of the neutron detector and of the fission chamber materials were studied separately. The biggest distortion to the $^{239}$Pu(n,f) PFN spectra comes from neutron scattering on the neutron detector materials. It is of about $1.5\%$ and is observed at the opening of the second-chance fission for an outgoing neutron energy below $250\,$keV. In the other region of the spectrum and for other incident neutron energies the distortion is negligible. A negligible distortion is also introduced by the presence of the fission chamber, as well as by two not working channels of the chamber itself (see Sec. \ref{sec:fission chamber}). 
 
To validate the assumption that the $^{252}$Cf PFNS is sufficiently close to the $^{239}$Pu, 
 the possible systematic bias introduced by the use of the $^{252}$Cf(s.f.) PFN spectra to determine the detector efficiencies were investigated. As an example, the ratios between the $^{239}$Pu(n,f) PFN spectrum determined as in the data analysis and the PFN spectrum used as input in the simulation are plotted in Fig.\ref{fig:simu pierre} for $0.87$ and $7.38\,$MeV incident neutron energies. The dashed line has the purpose of guiding the eyes.  A very good agreement is found for all the outgoing neutron energies for $0.87\,$MeV incident neutron energy, and, more in general, for incident neutron energies significantly different from the opening of the $2$-th chance fission. Indeed it is around this energy that a significant distortion is observed for outgoing-neutron energies below $1\,$MeV.
  \begin{figure}[t!]
\centering
\includegraphics[width=0.95\columnwidth,clip]{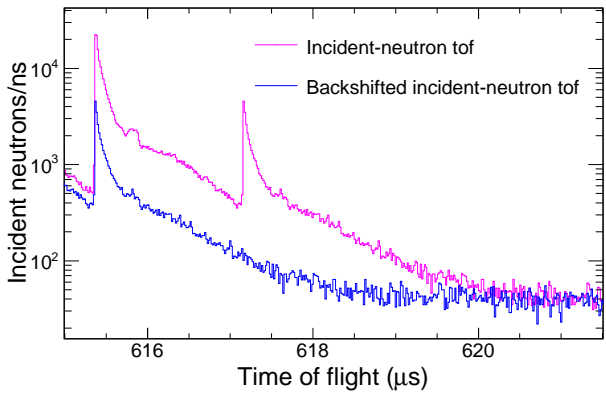}
\caption{(Color online) Incident-neutron time of flight from \cite{keegan}. The pink spectrum represents the last two micropulses of a macropulse whereas the blue one represents the last micropulse shifted by $1.8\,\mu$s. }
\label{fig:wa analysis} 
\end{figure}
 These results support the use of the $^{252}$Cf PFNS to account for the presence of surrounding materials and to determine the detector efficiencies, and indicate that it does not bias the experimental results. 
%
%
%
%
\subsection{Study of wrap-around background}\label{sec:wrap}
As discussed in Sec. \ref{sec:experimental setup}, due to the short micropulse spacing ($1.8\,\mu$s) and the long neutron flight-path ($\sim20\,$m), neutron time-of-flight distributions are affected by the so-called wrap-around background. Indeed slow neutrons ($\lesssim 680\,$keV) remaining after the polyethylene absorber reach the fission target at the same time or after the fastest neutrons of the following micropulse. Fission events induced by high-energy neutrons and low-energy neutron background can not be discriminated on time-of-flight basis. The fraction of wrap-around background in each incident neutron energy bin was estimated using data collected with a different DAQ, where the time of flight of neutrons belonging to the same macropulse are all stored \cite{keegan}. 
This allows one to inspect seperately fission events associated to each micropulse in a macropulse.
The last micropulse was considered. Indeed in this case, being two macropulses separated by about $625\,\mu$s, the slow neutrons do not overlap with the fastest neutrons of the following micropulse, and their time of flight can be cleanly characterized. This is clearly illustrated in Fig. \ref{fig:wa analysis} with the time spectrum of the incident neutron, in pink, which represents the last two micropulses of a macropulse.
%
%
%
\begin{figure}[t!]
\centering
\includegraphics[width=0.95\columnwidth,clip]{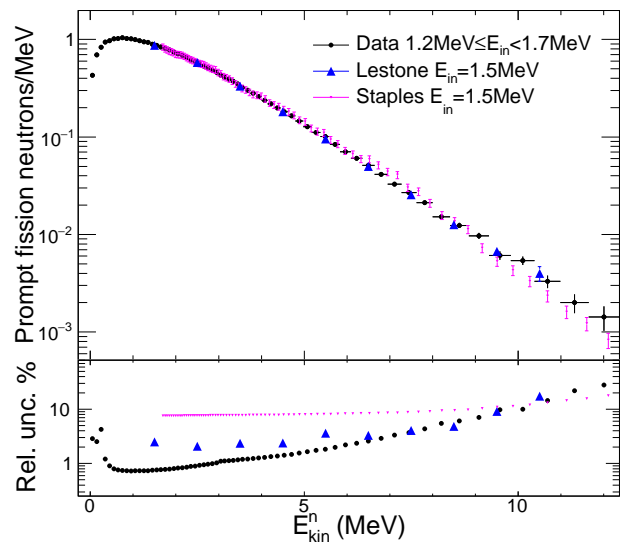} 
\caption{(Color online) Prompt fission neutron spectrum for an incident neutron energy ranging from $1.2$ to $1.7\,$MeV (black squares) compared to data from Lestone et al. \cite{lestone2014} (blue triangles) and from Staples et al. \cite{staples1995} (magenta triangles), measured at $1.5\,$MeV. In the lower panel the uncertainties of the three measurements are compared.}
\label{fig:pfns cfr lestone} 
\end{figure}
 \begin{figure*}[th!]
\centering
\includegraphics[width=1.95\columnwidth,clip]{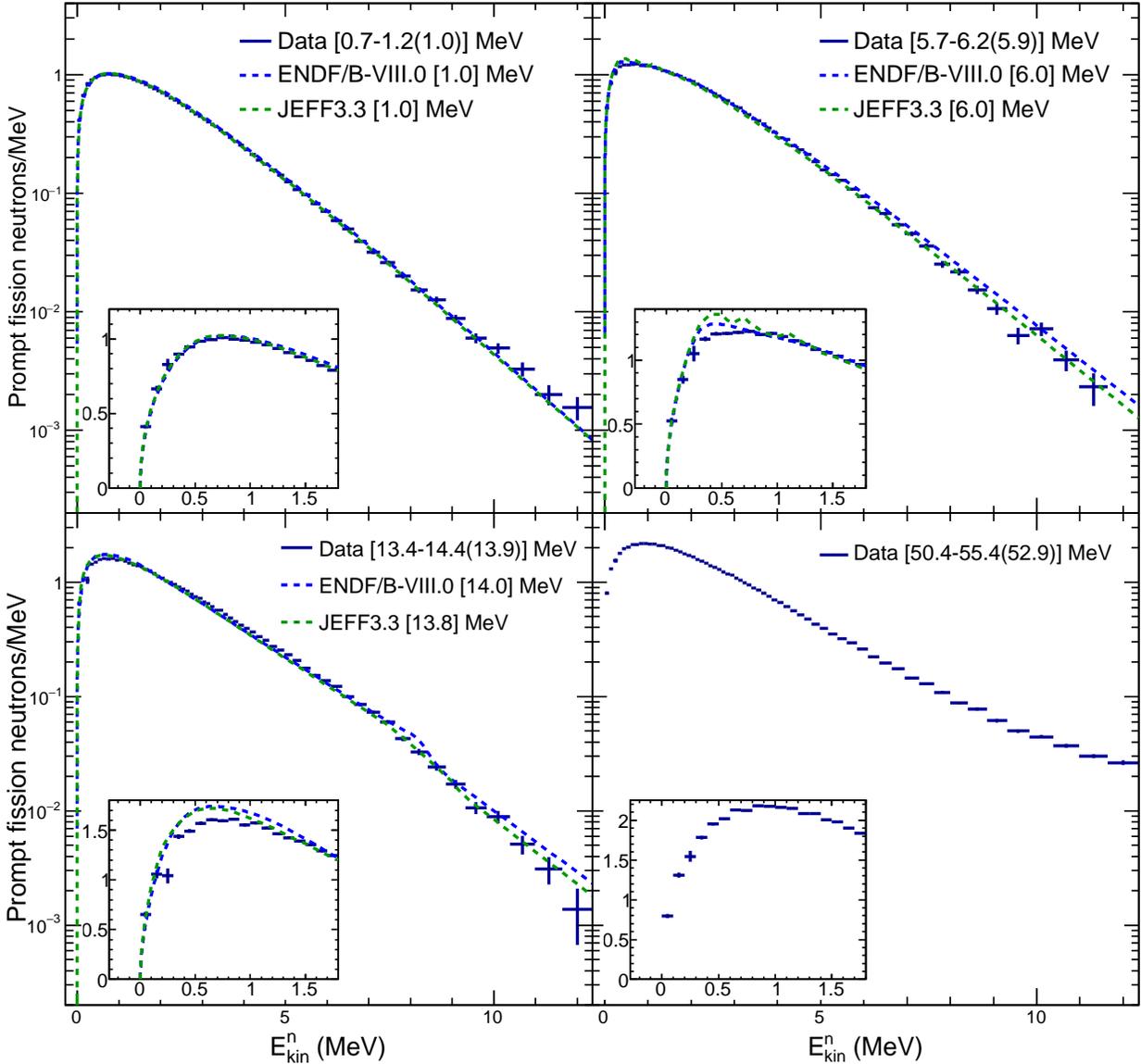}
\caption{(Color online) Prompt fission neutron spectrum for four beam energy ranges, indicated among square brackets, together with the mean incident neutrons kinetic energy of each bin. Experimental data (crosses) are compared to the
 ENDF/B-VIII.0 (dotted blue line) and JEFF 3.3 (dashed green line) evaluations. A zoom of the low energy part of the spectra is shown in the inserts.}
\label{fig:pfns for different ebin} 
\end{figure*}
 A good agreement was found between the incident-neutron time of flight measured in our experiment and the time of flight of incident neutrons belonging to the last micropulse measured in \cite{keegan}, with the appropriate normalization.
Hence these events can be use to determine the proportion of wrapped neutrons in the other micropulses.
  The time-of-fligh spectrum of the last micropulse was backshifted of $1.8\,\mu s$ (blue line in Fig. \ref{fig:wa analysis}). The backshifted-to-last micropulse time-of-flight ratio gives the fraction of slow neutron background present in each incident neutron energy bin.
The wrap-around contribution varies from a maximum of about $10\%$ to  about $3\%$ below $20\,$MeV and above $200\,$MeV, respectively. The associated relative uncertainty varies from $10\%$ to few $\%$ as the incident neutron energy increases, due to the statistics available in the used data.\\
It is worth recalling that slow neutrons giving rise to the wrap-around effect have energies $\lesssim 680\,$keV. The PFN spectra for these incident neutron energies were not measured during the experiment. However, under the reasonable assumption that PFN spectra for incident neutron energies below $1\,$MeV vary slowly, the PFN spectrum measured for an incident neutron energy ranging from $0.7$ to $1.2\,$MeV was used to correct PFN spectra for higher incident energies. It is worth noting that this introduces a correlation between all the PFN spectra. The uncertainties on the $0.7$ to $1.2\,$MeV incident energy PFN spectrum as well as on the wrap-around fraction were propagated to obtain wrap-around-corrected PFN spectra for all the incident energy bins.
In the following we will refer to these wrap-around corrected spectra as ``spectra'' and they will be used to calculate the quantities of interest, unless differently specified.
\section{Results}\label{sec:recults}
\subsection{PFNS}\label{sec:Results:PFNS}
\begin{figure}[h!]
\centering
\includegraphics[width=0.95\columnwidth,clip]{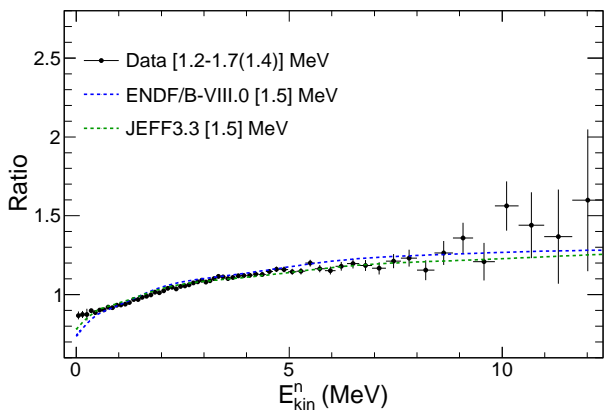}
\caption{(Color online) Ratio of PFNS to a Maxwellian distribution with $T=1.32\,$MeV for a mean incident energy of $1.44\,$MeV. Experimental data (crosses) are compared to the ENDF/B-VIII.0 (dotted blue line) and JEFF 3.3 (dashed green line) evaluations.} 
\label{fig:ratio to maxwell whole} 
\end{figure}
The PFNS measured for an average incident-neutron energy of $1.44\,$MeV is presented in Fig. \ref{fig:pfns cfr lestone} and compared to data from Lestone et al. \cite{lestone2014} and Staples et al. \cite{staples1995} for $1.5\,$MeV incident energy. The ratio of the PFN spectrum to a Maxwellian distribution with temperature $T=1.32\,$MeV is provided as supplemental material (Fig.\ref{fig:ratio to maxwell lestone staples}).
Our data are in very good agreement within the error bars with data from Lestone on the whole outgoing-energy range.
It should be noted that results from Lestone were obtained from underground nuclear explosion measurements, with a much higher neutron flux on the studied sample. A very good agreement is found also with data from Staples, with small discrepancies above about $9\,$MeV, due to  a change in the slope around $6\,$MeV, which leads to a slightly softer spectrum. Similar conclusions can be drawn when comparing the spectrum measured at $2.5\,$MeV, provided as supplemental material, with the one measured by Staples.

In the lower panel of Fig. \ref{fig:pfns cfr lestone} the relative uncertainty on the spectrum is presented (black circles). This is a typical example of the attained accuracy in the measurement for the smallest incident energy bins ($0.5\,$MeV bin width): between $0.3$ and $3\,$MeV the uncertainty is smaller than $1\%$,   it stays below $2\%$ up to $6\,$MeV and it is less than $10\%$ up to about $11\,$MeV.
These values have to be compared to the uncertainties obtained by Lestone, which vary from $2$ to $10\%$ between $1.5$ and $10.5\,$MeV, and to those measured by Staples, which are around $10\%$ on the whole outgoing-energy range.
\begin{figure*}[t]
\centering
\includegraphics[width=1.80\columnwidth,clip]{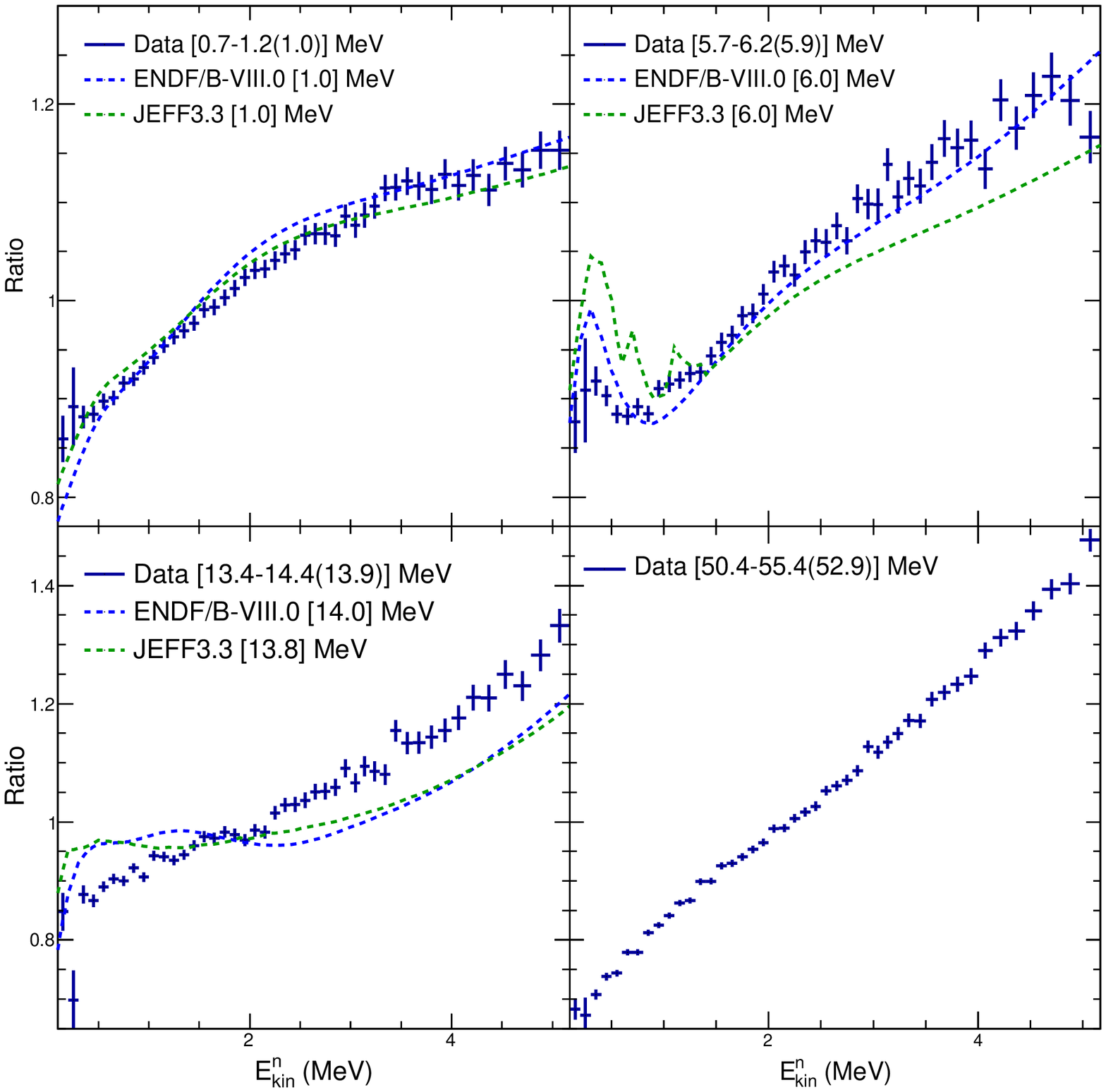}
\caption{(Color online) Zoom below $5\,$MeV neutron kinetic energy of the ratio of PFNS to a Maxwellian-type distribution with $T=1.32\,$MeV for four beam energies ranges. Beam energies are indicated in square brackets, together with the mean incident neutrons kinetic energy of each bin. Experimental data (crosses) are compared to ENDF/B-VIII.0 (dotted blue line) and JEFF 3.3 (dashed green line) evaluations.}
\label{fig:ratio to maxwell} 
\end{figure*}
\begin{figure*}[ht!]
\centering
\includegraphics[width=1.80\columnwidth,clip]{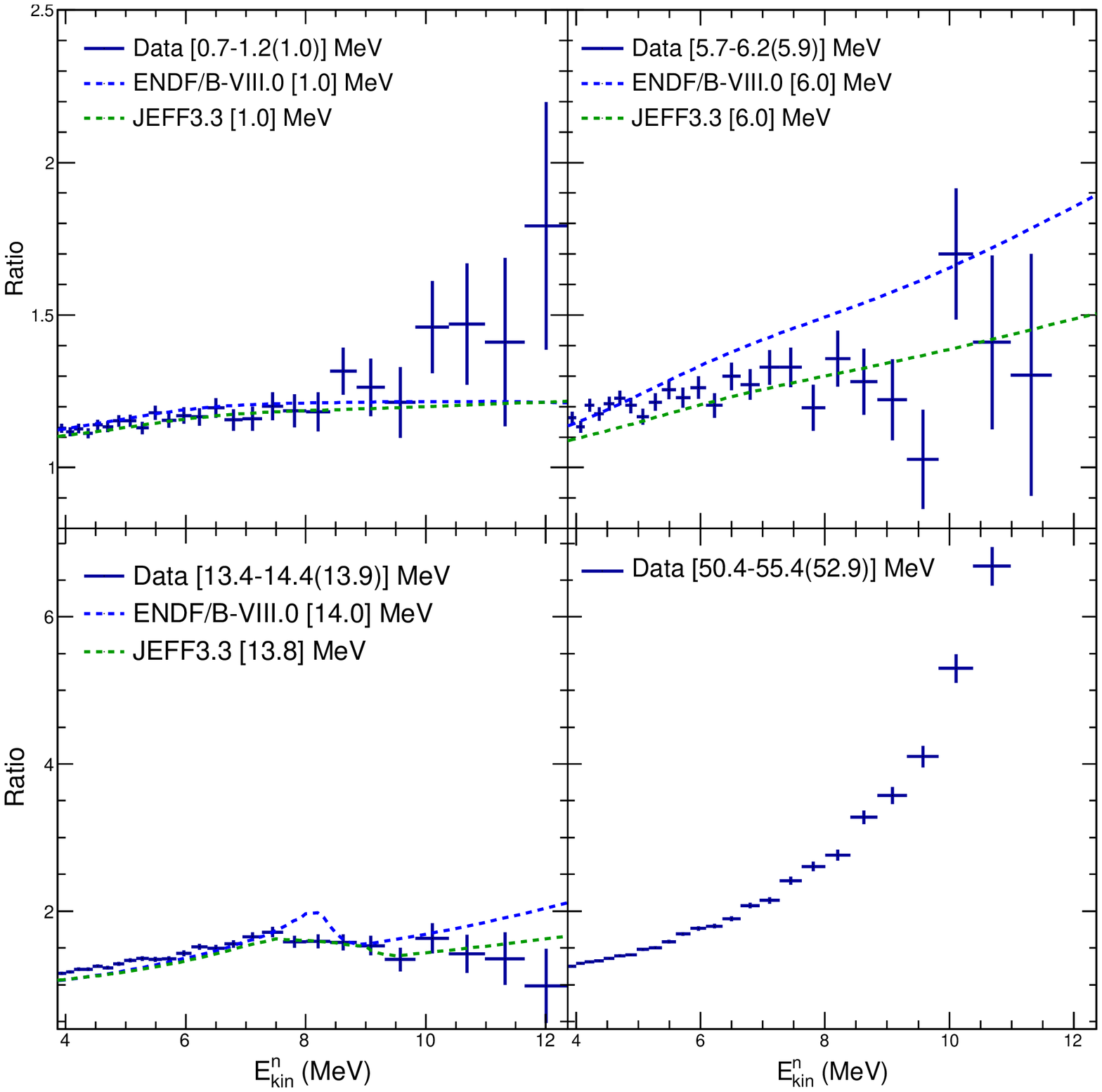}
\caption{(Color online) Zoom above $4\,$MeV neutron kinetic energy of the ratio of PFNS to a Maxwellian-type distribution with $T=1.32\,$MeV for four beam energies ranges. As Fig. \ref{fig:ratio to maxwell}.}
\label{fig:ratio to maxwell zoom 2} 
\end{figure*}
Figure \ref{fig:pfns for different ebin} shows spectra for four incident neutron energy bins, compared to the ENDF/B-VIII.0 \cite{endf8,endf8_pfns} and JEFF3.3 \cite{JEFF33} evaluations. Spectra for the other energy bins are provided as supplemental material. All the spectra are normalized to the integral of the PFNS on the whole energy range (i.e., to the experimental neutron multiplicity). 
Overall, a fair agreement is observed between the data and the evaluations. Slight discrepancies are found for the low- ($<1\,$MeV) and  high-energy tails ($>9\,$MeV) of the spectra, depending on the incident neutron energy.
We concentrate here on outgoing-neutron kinetic energies above $3\,$MeV.
For spectra up to 
the opening of the second-chance fission ($\simeq6\,$MeV), a rather good agreement is observed between the data and the evaluations. 
For energies above the second 
chance fission, a better agreement is observed at high kinetic energy with the \jeff evaluation, while \ENDF  predicts  harder spectrum (a higher number of high-energy fission neutrons). Data indicate that both the evaluations predict harder spectra for energies above $18\,$MeV.
%

Figures \ref{fig:ratio to maxwell whole}, \ref{fig:ratio to maxwell} and \ref{fig:ratio to maxwell zoom 2} show the ratio of PFN spectra to a Maxwellian distribution with temperature $T=1.32\,$MeV for five incident-neutron energy bins, compared to  \ENDF and \jeff evaluations. In Fig. \ref{fig:ratio to maxwell} only emitted neutrons with kinetic energies below $5\,$MeV are shown, to focus the attention on the low-energy part of the spectra. On the contrary Fig. \ref{fig:ratio to maxwell zoom 2} focuses on the high-energy part of the spectra. The spectra, as well as the evaluations, were normalized to the integral of the PFN spectra on the whole energy range.
For spectra up to $3.4\,$MeV incident-neutron energy, predictions seem to underestimate the amount of fission neutrons below $500\,$keV, while a better agreement is found for higher incident energies. The situation changes for energies around the opening of the second-, third and forth-chance fission ($6$, $14$ and $24\,$MeV, respectively), where discrepancies are observed for outgoing energies below $1.5\,$MeV.

\subsection{PFN mean kinetic energy}
\begin{figure}[h!]
\centering
\includegraphics[width=0.95\columnwidth,clip]{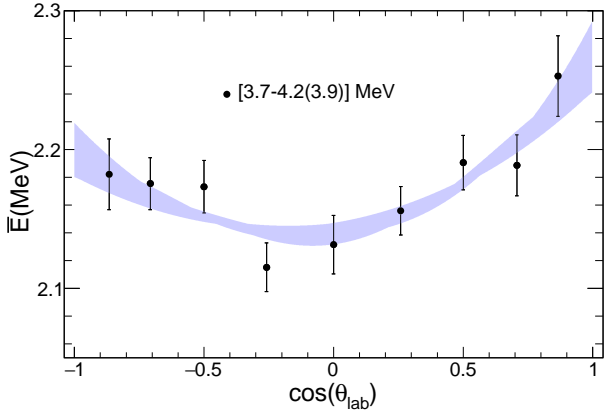}
\caption{Angular dependence of \emean$(\theta_{lab},E_{in})$ for an incident neutron energy ranging from $3.7$ to $4.2\,$MeV. The shaded area is the envelop of the best polynomial fits to the experimental data (see text). }
\label{fig:emean vs theta} 
\end{figure}

The high statistics  accumulated during the experiment and the Chi-Nu segmentation allows us to obtain a  precise shape of the PFNS as a function of the neutron emission angle with respect to the beam direction ($\theta_{lab}$) for each incident-neutron energy range studied. Nine spectra, one for each measured $\theta_{lab}$, were therefore obtained by combining the spectra obtained from the six detectors at the considered angle. 
The average neutron kinetic energy, \emean, was computed from the spectra for every $\theta_{lab}$ and incident energy bin. No threshold was applied.
An example of angular dependence of \emean, \emean$(\theta_{lab},E_{in})$, is plotted in Fig. \ref{fig:emean vs theta} for an incident neutron energy ranging from $3.7$ to $4.2\,$MeV. 
%
%
The total uncertainty on \emean $(\theta,E_{in})$ is close to $0.3\,$\%. It includes the statistical uncertainties on the \pu and $^{252}$Cf measurements, on their background measurements and on the wrap-around correction. The systematic uncertainty encompasses the uncertainty of the evaluated $^{252}$Cf(s.f.) PFN spectra. We remind that the associated correlation matrix was not propagated. The angular distributions exhibit two main characteristics: first, they are not isotropic, even at low incident energies, with a \emean$(0^{\circ})/$\emean$(90^{\circ})$ of about $1.04$ below $10\,$MeV. 
The anisotropy arises from the fact that PFN carry angular momentum from fully accelerated fission fragments. Information on fission fragments angular momentum at scission could therefore be extracted by model comparison, which is beyond the scope of this paper.
 Second, angular distributions are characterized by a certain degree of forward/backward asymmetry which increases with the incident-neutron energy, reflecting the increase in the  kinematical boost and  pre-equilibrium emission.

%

The experimentally measured mean kinetic energy 
angular distributions 
were also needed  in the determination of the  average neutron \emean$(E_{in})$ 
for each incident neutron energy range. 
They 
were  fitted with polynomial functions and  
\emean$(E_{in})$ 
was taken as the sum of the experimental values and the values deduced from the fitted distribution, for those angles that were not covered by detectors during the experiment. 
 Several different fitting functions were tested to estimate the systematic uncertainty on \emean \, due to the arbitrary choice of the functions. It amounts to about $0.2\%$ and was calculated as the root-mean-square of the \emean values obtained with the different fitting functions for each incident neutron energy. The envelop of the different fitting functions used is shown as shaded area in Fig.\ref{fig:emean vs theta}. The total uncertainty is  estimated to be between $0.2\%$ and $0.5\%$ depending on the incident neutron energy bin.

\begin{figure}[bh!]
\centering
\includegraphics[width=0.95\columnwidth,clip]{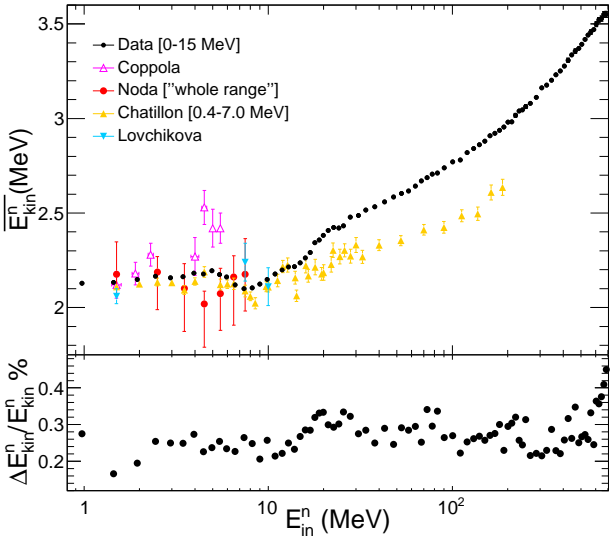}
\caption{(Color online) Measured mean kinetic energy \emean$(E_{in})$ of PFN as a function of incident neutron energy compared to data from previous experiments \cite{Coppola1970,noda2011,chatillon,Lovchikova1996} (panel a). Among brackets is the range over which  \emean$(E_{in})$ was calculated, when specified in the publications. Relative uncertainty on \emean$(E_{in})$  a function of incident neutron energy (panel b). } 
\label{fig:emean vs ein donnees} 
\end{figure}

%
%
%
The obtained values are plotted in Fig. \ref{fig:emean vs ein donnees} 
as a function of the kinetic energy of the incoming neutron. All available other experimental results are also plotted on the same graph \cite{Coppola1970,noda2011,chatillon,Lovchikova1996}.
The mean kinetic energy of the emitted neutrons increases rather slowly over the full energy range, spanning from $2.1$ to $3.5\,$MeV.
Above $20\,$MeV, we observe a slower increase of the kinetic energy, while 
fluctuations are observed below this energy. 
A clear dip of about $100\,$keV depth is observed around $6\,$ MeV, and described with 8 experimental data points between $5$ and $11\,$MeV. Smaller second and third dips are 
observed around $14$ and $24\,$MeV, as reported also in \cite{chatillon}. The presence of these dips is due to pre-fission neutrons emitted at the opening of the second- (third- and fourth-) chance fission. Indeed pre-fission neutrons, evaporated from the compound nucleus before fissioning, have smaller kinetic energy than fission neutrons, since they do not profit from the kinematical boost from fission fragments. 

A good agreement with the most recent data \cite{chatillon} is found up to $14\,$MeV. At higher energy, where pre-equilibrium reactions set in, the discrepancy is explained by the fact that the setup used in Ref. \cite{chatillon} is most sensitive to lower energy evaporation neutrons and not to the higher energy neutrons emitted in the forward direction. 
 A fair agreement is observed also with data from Refs.\cite{Lovchikova1996,noda2011}, which present large error bars. Our data are not consistent with data from Ref. \cite{Coppola1970}.

\begin{figure}[t!]
\centering
\includegraphics[width=0.95\columnwidth,clip]{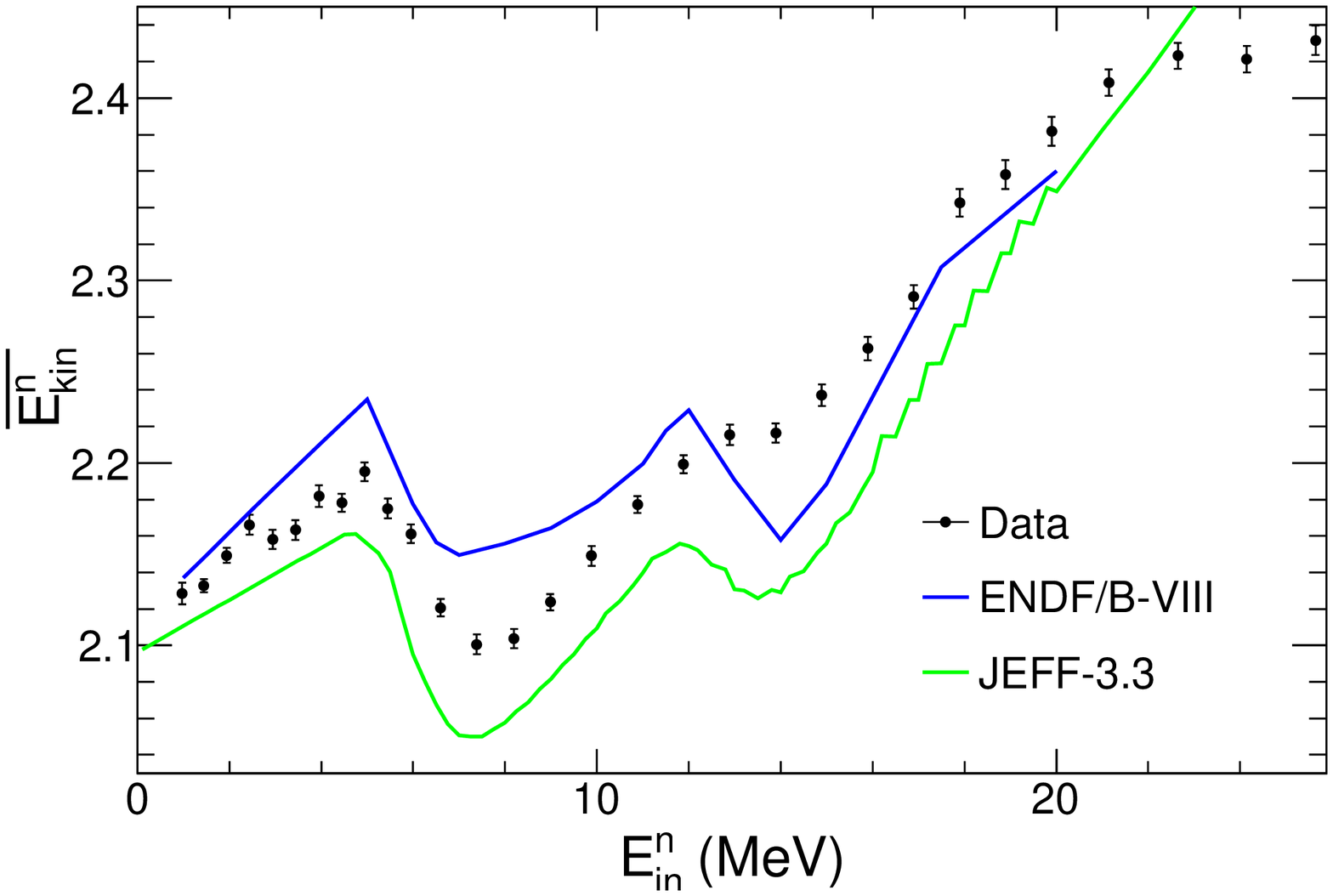}
\caption{Measured mean kinetic energy \emean$(E_{in})$ compared to 
\ENDF and \jeff evaluations. } 
\label{fig:emean vs ein gef} 
\end{figure}

Figure \ref{fig:emean vs ein gef} shows the comparison of our data to 
\ENDF and \jeff evaluations. 
Below the opening of the second-chance fission 
the \jeff evaluation reproduce the trend of \emean, although the values are slightly underestimated. On the contrary \ENDF indicates a steeper increase of \emean$\,$ with increasing the neutron energy, which is not observed in the data.
Above $5\,$MeV  JEFF3.3, and to a lesser extent \ENDF,  reproduce the dip depth, although they are not consistent with the experimental values. Both  evaluations indicate the presence of a second dip around $12\,$MeV, whose depth is much less pronounced in the experimental data than in the evaluations. It appears that all 
evaluations either underestimate (\jeff) or overestimate the second and third chance fission probability in \pu($n,f$). These features notably impact  the PFNS and the mean neutron energy.  None of the evaluations is in agreement with the measured data.  
%
%
%
%
%
\section{Conclusions}\label{sec:conclusions}
Prompt fission neutron spectra from $0.2$ to $10-12\,$MeV outgoing-neutron energy were measured with respect to $^{252}$Cf spontaneous fission with the Chi-Nu liquid scintillator array for incident neutron energies from $1$ to $700\,$ MeV. The obtained final uncertainties are well below $2\%$ for outgoing-neutron energies up to $6\,$MeV, and lower than the ones from all previous measurements.\\
In general, the data agree well with theoretical predictions based on the Los Alamos model, \ENDF and JEFF3.3, from $1$ to about $10\,$MeV outgoing-neutron energies. 
Below about $3\,$MeV incident energy, evaluations underestimate 
the number of low-energy neutrons ($<500\,$keV) and significant discrepancies are observed below outgoing energies of $1.5\,$MeV around the opening of the second-, third- and forth-chance fission.
Above  $6\,$MeV incident energy, a better agreement is found at high kinetic energy with the \jeff evaluation.


The mean energy of the PFNS was measured with a
total uncertainty better than $0.5\%$ for all incident neutron energies and shows a continuous increase with the beam energy. Signatures of the opening of the second- third- and fourth-chance fissions can be recognized around $6$, $14$ and $24\,$MeV. A good agreement is found with previous data from Ref. \cite{chatillon} up to the onset of pre-equilibrium reactions, around $10-12\,$MeV. The shape, but not the measured values, is better reproduced by the \jeff evaluation.

The obtained accuracies were made possible thanks to a newly developed fission chamber with a detection efficiency close to $100\%$, its associated electronics, the high statistics collected and the Chi-Nu segmentation and neutron/$\gamma$ discrimination capabilities.
Given the success of the present experiment, further measurements should be performed to provide more accurate PFNS data than  those available at present on other relevant actinides. \\
%
\newpage
\section*{Acknowledgments}
We wish to acknowledge A.~Moens, G.~Sibbens and D.~Vanleeuw from the JRC-Geel target preparation laboratory for providing $^{239}$Pu samples and assisting their mounting in the fission chamber.
We also wish to acknowledge the support of  E.~Bond from LANL C-NR for providing the $^{252}$Cf sample.
This work was performed under the auspices of a cooperation agreement between CEA/DAM and DOE/NNSA on fundamental sciences and benefited from the use of the LANSCE accelerator facility.
 The work at Los Alamos was performed under the auspices of the U.S. Department of Energy by Los Alamos National Laboratory under Contract DE-AC52-06NA25396.

%

\bibliographystyle{unsrt}
\bibliography{bibliogr}

\begin{thebibliography}{10}

\bibitem{peneliau2014}
Y.~Peneliau, O.~Litaize, P.~Archier, and C.~De~Saint Jean.
\newblock {\em Nucl. Data Sheets}, 118:459 -- 462, 2014.

\bibitem{capote2008}
R.~Capote, V.~Maslov, E.~Bauge, T.~Ohsawa, A.~Vorobyev, M.B. Chadwick, and
  S.~Oberstedt.
\newblock Report consultants' meeting on prompt fission neutron spectra.
\newblock Technical report, Vienna, INDC(NDS) 0541, 2008.

\bibitem{neudecker2016}
D.~Neudecker, T.N. Taddeucci, R.C. Haight, H.Y. Lee, M.C. White, and M.E.
  Rising.
\newblock {\em Nucl. Data Sheets}, 131:289 -- 318, 2016.

\bibitem{nefedov}
V.~N. Nefedov, B.~I. Starostov, and A.~A. Boytsov.
\newblock Precision measurements of $^{252}\mathrm{Cf}$, $^{233}\mathrm{U}$,
  $^{235}\mathrm{U}$ and $^{239}\mathrm{Pu}$ prompt fission neutron spectra
  ({PFNS}) in the energy range 0.04-5 {M}e{V}.
\newblock In {\em IAEA Vienna Report INDC(CCP)-0457 (2014), translation into
  English from: Proceedings of the All-Union Conf. on Neutron Physics, Kiev,
  2-6 Oct. 1983 2, 285 (1983) (in Russian), EXFOR-No. 40871}.

\bibitem{boytsov}
A.~A. Boytsov, A.~F. Semenov, and B.~I. Starostov.
\newblock Relative measurements of $^{233}\mathrm{U}$+$\mathrm{n}_{th}$,
  $^{235}\mathrm{U}$+$\mathrm{n}_{th}$ and
  $^{239}\mathrm{Pu}$+$\mathrm{n}_{th}$ prompt fission neutron spectra ({PFNS})
  in the energy range 0.01-5 {M}e{V}.
\newblock In {\em IAEA Vienna Report INDC(CCP)-0459 (2014), translation into
  English from: Proceedings of the All-Union Conf. on Neutron Physics, Kiev,
  2-6 Oct. 1983 2, 294 (1983) (in Russian), EXFOR-No. 40873.}

\bibitem{starostov}
B.~I. Starostov, V.~N. Nefedov, and A.~A. Boytsov.
\newblock Precision measurements of $^{252}\mathrm{C}$f,
  $^{233}\mathrm{U}$+$\mathrm{n}_{th}$, $^{235}\mathrm{U}$+$\mathrm{n}_{th}$
  and $^{239}\mathrm{Pu}$+$\mathrm{n}_{th}$ prompt fission neutron spectra
  ({PFNS}) in the energy range 2-11 {MeV}.
\newblock In {\em IAEA Vienna Report INDC(CCP)-0458 (2014), translation into
  English from: Proceedings of the All-Union Conf. on Neutron Physics, Kiev,
  2-6 Oct. 1983 2, 290 (1983) (in Russian), EXFOR-No. 40872}.

\bibitem{starostov89}
B.~I. Starostov, V.~N. Nefedov, and A.~A. Bojtsov.
\newblock Prompt neutron spectra from fission of $^{233}\mathrm{U}$,
  $^{235}\mathrm{U}$ and $^{239}\mathrm{Pu}$ by thermal neutrons and from
  spontaneous fission of $^{252}\mathrm{C}$f in the 0.01-12 {MeV} energy range.
\newblock In {\em IAEA Vienna Report INDC(CCP)-293/L, p.19-32 (1989),
  translation into English from: Yadernye Konstanty 1985, 16 (1985) (in
  Russian), EXFOR-No. 40930}.

\bibitem{lajtai}
A.~Lajtai, J.~Kecskem\'eti, J.~S\'af\'ar, P.~P. Dyachenko, and V.~M. Piksaikin.
\newblock {Energy Spectrum Measurements of Neutrons for Energies 30 keV-4 MeV
  from Thermal Fission of Main Fuel Elements}.
\newblock In {\em Proceedings of the Conf. on Nuclear Data for Basic and
  Applied Sciences, Santa Fe 1985 1, 613 (1985), EXFOR-No. 30704.}

\bibitem{staples1995}
P.~Staples, J.~J. Egan, G.~H.~R. Kegel, A.~Mittler, and M.~L. Woodring.
\newblock {\em Nucl. Phys. A}, 591:41 -- 60, 1995.
\newblock EXFOR-No. 13982.

\bibitem{knitter75}
H.~H. Knitter.
\newblock {\em Atomkernenergie}, 26:76, 1975.
\newblock EXFOR-No. 20576.

\bibitem{lestone2014}
J.~P. Lestone and E.~F. Shores.
\newblock {\em Nucl. Data Sheets}, 119:213 -- 216, 2014.

\bibitem{chatillon}
A.~Chatillon, G.~B\'elier, T.~Granier, B.~Laurent, B.~Morillon, J.~Taieb, R.~C.
  Haight, M.~Devlin, R.~O. Nelson, S.~Noda, and J.~M. O'Donnell.
\newblock {\em Phys. Rev. C}, 89:014611, 2014.

\bibitem{noda2011}
S.~Noda, R.~C. Haight, R.~O. Nelson, M.~Devlin, J.~M. O'Donnell, A.~Chatillon,
  T.~Granier, G.~B\'elier, J.~Taieb, T.~Kawano, and P.~Talou.
\newblock {\em Phys. Rev. C}, 83:034604, 2011.

\bibitem{capote2016}
R.~Capote, Y.-J. Chen, F.-J. Hambsch, N.V. Kornilov, J.~P. Lestone, O.~Litaize,
  B.~Morillon, D.~Neudecker, S.~Oberstedt, T.~Ohsawa, N.~Otuka, V.G. Pronyaev,
  A.~Saxena, O.~Serot, O.A. Shcherbakov, N.-C. Shu, D.~L. Smith, P.~Talou,
  A.~Trkov, A.~C. Tudora, R.~Vogt, and A.~S. Vorobyev.
\newblock {\em Nucl. Data Sheets}, 131:1 -- 106, 2016.

\bibitem{neudecker2018}
D.~Neudecker, P.~Talou, T.~Kawano, A.C. Kahler, M.C. White, T.N. Taddeucci,
  R.C. Haight, B.~Kiedrowski, J.M. O'Donnell, J.A. Gomez, K.J. Kelly,
  M.~Devlin, and M.E. Rising.
\newblock {\em Nucl. Data Sheets}, 148:293, 2018.

\bibitem{madland_LosAlamosModel1}
D.~G. Madland and J.~Rayford Nix.
\newblock {\em Nucl. Sci. Eng.}, 81:213--271, 1982.

\bibitem{chadwick2011}
M.B. Chadwick, M.~Herman, P.~Oblo\v{z}insk\'{y}, M.~E. Dunn, Y.~Danon, A.~C.
  Kahler, D.~L. Smith, B.~Pritychenko, G.~Arbanas, R.~Arcilla, R.~Brewer, D.~A.
  Brown, R.~Capote, A.~D. Carlson, Y.S. Cho, H.~Derrien, K.~Guber, G.M. Hale,
  S.~Hoblit, S.~Holloway, T.D. Johnson, T.~Kawano, B.C. Kiedrowski, H.~Kim,
  S.~Kunieda, N.M. Larson, L.~Leal, J.~P. Lestone, R.~C. Little, E.A.
  McCutchan, R.E. MacFarlane, M.~MacInnes, C.M. Mattoon, R.D. McKnight, S.F.
  Mughabghab, G.P.A. Nobre, G.~Palmiotti, A.~Palumbo, M.T. Pigni, V.G.
  Pronyaev, R.O. Sayer, A.A. Sonzogni, N.C. Summers, P.~Talou, I.J. Thompson,
  A.~Trkov, R.L. Vogt, S.C. van~der Marck, A.~Wallner, M.C. White, D.~Wiarda,
  and P.G. Young.
\newblock {\em Nucl. Data Sheets}, 112:2887 -- 2996, 2011.

\bibitem{maslov2011}
V.~M. Maslov, N.~A. Tetereva, V.~G. Pronyaev, A.~B. Kagalenko, K.~I. Zolotarev,
  R.~Capote, T.~Granier, B.~Morillon, F.-J. Hambsch, and J.-C. Sublet.
\newblock {\em J. Korean Phy. Soc.}, 59:1337, 2011.

\bibitem{talou2011}
P.~Talou, B.~Becker, T.~Kawano, M.~B. Chadwick, and Y.~Danon.
\newblock {\em Phys. Rev. C}, 83:064612, 2011.

\bibitem{freya}
J.~Randrup and R.~Vogt.
\newblock {\em Phys. Rev. C}, 80:024601, 2009.

\bibitem{fifrelin}
O.~Litaize and O.~Serot.
\newblock {\em Phys. Rev. C}, 82:054616, 2010.

\bibitem{madland_LosAlamosModel2}
D.~G. Madland.
\newblock {\em Acta Phys. Hungaria New Ser. Heavy Ion Phys.}, 10:231, 1999.

\bibitem{karlheinz2010}
K.-H. Schmidt and B.~Jurado.
\newblock {\em Phys. Rev. Lett.}, 104:212501, 2010.

\bibitem{karlheinz2011}
K.-H. Schmidt and B.~Jurado.
\newblock {\em Phys. Rev. C}, 83:014607, 2011.

\bibitem{karlheinz2011_2}
K.-H. Schmidt and B.~Jurado.
\newblock {\em Phys. Rev. C}, 83:061601(R), 2011.

\bibitem{ethvignot}
T.~Ethvignot, M.~Devlin, R.~Drosg, T.~Granier, R.C. Haight, B.~Morillon, R.O.
  Nelson, J.M. O'Donnell, and D.~Rochman.
\newblock {\em Phys. Lett. B}, 575:221 -- 228, 2003.

\bibitem{taiebProc}
J.~Taieb, T.~Granier, T.~Ethvignot, M.~Devlin, R.C. Haight, R.O. Nelson, J.M.
  O'Donnell, and D.~Rochman.
\newblock In {\em Proceedings of the International Conference on Nuclear Data
  for Science and Technology}, page 429, 2007.

\bibitem{keegan_Pu}
K.~J. Kelly~et al.
\newblock to be submitted to Phys. Rev. C.

\bibitem{devlin2018}
M.~Devlin, J.A. Gomez, K.J. Kelly, R.C. Haight, J.M. O'Donnell, T.N. Taddeucci,
  H.Y. Lee, S.M. Mosby, B.A. Perdue, N.~Fotiades, J.L. Ullmann, C.Y. Wu,
  B.~Bucher, M.Q. Buckner, R.A. Henderson, D.~Neudecker, M.C. White, P.~Talou,
  M.E. Rising, and C.J. Solomon.
\newblock {\em Nuclear Data Sheets}, 148:322, 2018.

\bibitem{wnr1}
P.~W. Lisowski, C.~D. Bowman, G.~J. Russell, and S.~A. Wender.
\newblock {\em Nucl. Sci. Eng.}, 106:208, 1990.

\bibitem{wnr}
P.~W. Lisowski and K.~F. Schoenberg.
\newblock {\em Nucl. Instrum. Methods Phys. Res., Sect. A}, 562:910 -- 914,
  2006.

\bibitem{taieb}
J.~Taieb, B.~Laurent, G.~B\'elier, A.~Sardet, and C.~Varignon.
\newblock {\em Nucl. Instrum. Methods Phys. Res., Sect. A}, 833:1 -- 7, 2016.
\newblock Patent pending for the anode concept.

\bibitem{chinu}
R.~C. Haight, H.~Y. Lee, T.~N. Taddeucci, J.~M. O'Donnell, B.~A. Perdue,
  N.~Fotiades, M.~Devlin, J.~L. Ullmann, A.~Laptev, T.~Bredeweg, M.~Jandel,
  R.~O. Nelson, S.~A. Wender, M.~C. White, C.~Y. Wu, E.~Kwan, A.~Chyzh,
  R.~Henderson, and J.~Gostic.
\newblock {\em J. Inst.}, 7:C03028, 2012.

\bibitem{budtz}
C.~Budtz-J{\o}rgensen and H.-H. Knitter.
\newblock Investigation of fission layers for precise fission cross-section
  measurements with a gridded ionization chamber.
\newblock {\em Nuclear Science and Engineering}, 86(1):10--21, 1984.

\bibitem{pino2014}
F.~Pino, L.~Stevanato, D~Cester, G.~Nebbia, L.~Sajo-Bohus, and G.~Viesti.
\newblock {\em Appl. Radiat. Isot.}, 89:79 -- 84, 2014.

\bibitem{faster}
{FASTER}.
\newblock {LPC-C}aen \url{http://faster.in2p3.fr}, 2013.

\bibitem{polack}
J.~K. Polack, M.~Flaska, A.~Enqvist, C.~S. Sosa, C.~C. Lawrence, and S.~A.
  Pozzi.
\newblock {\em Nucl. Instrum. Methods Phys. Res., Sect. A}, 795:253 -- 267,
  2015.

\bibitem{carlson2009}
A.~D. Carlson, V.~G. Pronyaev, D.~L. Smith, N.~M. Larson, Zhenpeng Chen, G.~M.
  Hale, F.-J. Hambsch, E.~V. Gai, Soo-Youl Oh, S.~A. Badikov, T.~Kawano, H.~M.
  Hofmann, H.~Vonach, and S.~Tagesen.
\newblock {\em Nucl. Data Sheets}, 110:3215 -- 3324, 2009.

\bibitem{carlson2018}
A.~D. Carlson, V.~G. Pronyaev, R.~Capote, G.~M. Hale, Z.-P. Chen, I.~Duran,
  F.-J. Hambsch, S.~Kunieda, W.~Mannhart, B.~Marcinkevicius, R.~O. Nelson,
  D.~Neudecker, G.~Noguere, M.~Paris, S.~P. Simakov, P.~Schillebeeckx, D.~L.
  Smith, X.~Tao, A.~Trkov, A.~Wallner, and W.~Wang.
\newblock {\em Nucl. Data Sheets}, 148:143 -- 188, 2018.

\bibitem{watt1952}
B.~E. Watt.
\newblock {\em Phys. Rev.}, 87:1037--1041, 1952.

\bibitem{bloch1943}
F.~Bloch and H.~Staub.
\newblock {U.S. Atomic Energy Commission Document AECD-3158}.
\newblock (as quoted by Terrell in Phys. Rev. 113:527-541, 1959), 1943.

\bibitem{terrell1959}
J.~Terrell.
\newblock {\em Phys. Rev.}, 113:527--541, 1959.

\bibitem{GEF}
K.-H. Schmidt, B.~Jurado, C.~Amouroux, and C.~Schmitt.
\newblock {\em Nucl. Data Sheets}, 131:107 -- 221, 2016.

\bibitem{geant4}
J.~Allison, K.~Amako, J.~Apostolakis, P.~Arce, M.~Asai, T.~Aso, E.~Bagli,
  A.~Bagulya, S.~Banerjee, G.~Barrand, B.R. Beck, A.G. Bogdanov, D.~Brandt,
  J.M.C. Brown, H.~Burkhardt, Ph. Canal, D.~Cano-Ott, S.~Chauvie, K.~Cho,
  G.A.P. Cirrone, G.~Cooperman, M.A. Cort\'es-Giraldo, G.~Cosmo, G.~Cuttone,
  G.~Depaola, L.~Desorgher, X.~Dong, A.~Dotti, V.D. Elvira, G.~Folger,
  Z.~Francis, A.~Galoyan, L.~Garnier, M.~Gayer, K.L. Genser, V.M. Grichine,
  S.~Guatelli, P.~Gu\`eye, P.~Gumplinger, A.S. Howard,
  I.~H\v{r}ivn\'a\v{c}ov\'a, S.~Hwang, S.~Incerti, A.~Ivanchenko, V.N.
  Ivanchenko, F.W. Jones, S.Y. Jun, P.~Kaitaniemi, N.~Karakatsanis,
  M.~Karamitros, M.~Kelsey, A.~Kimura, T.~Koi, H.~Kurashige, A.~Lechner, S.B.
  Lee, F.~Longo, M.~Maire, D.~Mancusi, A.~Mantero, E.~Mendoza, B.~Morgan,
  K.~Murakami, T.~Nikitina, L.~Pandola, P.~Paprocki, J.~Perl, I.~Petrovi\'c,
  M.G. Pia, W.~Pokorski, J.M. Quesada, M.~Raine, M.A. Reis, A.~Ribon,
  A.~Risti\'c Fira, F.~Romano, G.~Russo, G.~Santin, T.~Sasaki, D.~Sawkey, J.I.
  Shin, I.I. Strakovsky, A.~Taborda, S.~Tanaka, B.~Tomé, T.~Toshito, H.N.
  Tran, P.R. Truscott, L.~Urban, V.~Uzhinsky, J.M. Verbeke, M.~Verderi, B.L.
  Wendt, H.~Wenzel, D.H. Wright, D.M. Wright, T.~Yamashita, J.~Yarba, and
  H.~Yoshida.
\newblock {\em Nuclear Instruments and Methods in Physics Research Section A:
  Accelerators, Spectrometers, Detectors and Associated Equipment}, 835:186 --
  225, 2016.

\bibitem{matta}
A.~Matta, P.~Morfouace, N.~de~S{\'{e}}r{\'{e}}ville, F.~Flavigny, M.~Labiche,
  and R.~Shearman.
\newblock {\em Jour. of Phys. G: Nucl. and Part. Phys.}, 43:045113, 2016.

\bibitem{kohley_menate}
Z.~Kohley, E.~Lunderberg, P.A. DeYoung, B.T. Roeder, T.~Baumann, G.~Christian,
  S.~Mosby, J.K. Smith, J.~Snyder, A.~Spyrou, and M.~Thoennessen.
\newblock {\em Nucl. Instr. and Meth. in Phys. Res. Sec. A}, 682:59, 2012.

\bibitem{keegan}
K.J. Kelly, J.A. Gomez, J.M. O'Donnell, M.~Devlin, R.C. Haight, T.N. Taddeucci,
  S.M. Mosby, H.Y. Lee, D.~Neudecker, M.C. White, C.Y. Wu, R.~Henderson,
  J.~Henderson, M.Q. Buckner, P.~Talou, N.~Fotiades, M.E. Rising, and C.J.
  Solomon.
\newblock In {\em Proceedings of the 20th Topical Meeting of the Radiation
  Protection and Shielding Division}.
\newblock LA-UR-18-28140 (2018).

\bibitem{endf8}
D.~A. Brown, M.~B. Chadwick, R.~Capote, A.~C. Kahler, A.~Trkov, M.~W. Herman,
  A.~A. Sonzogni, Y.~Danon, A.~D. Carlson, M.~Dunn, D.~L. Smith, G.~M. Hale,
  G.~Arbanas, R.~Arcilla, C.~R. Bates, B.~Beck, B.~Becker, F.~Brown, R.~J.
  Casperson, J.~Conlin, D.~E. Cullen, M.-A. Descalle, R.~Firestone, T.~Gaines,
  K.~H. Guber, A.~I. Hawari, J.~Holmes, T.~D. Johnson, T.~Kawano, B.~C.
  Kiedrowski, A.~J. Koning, S.~Kopecky, L.~Leal, J.~P. Lestone, C.~Lubitz,
  J.~I.~M\'{a}rquez Dami\'{a}n, C.~M. Mattoon, E.~A. McCutchan, S.~Mughabghab,
  P.~Navratil, D.~Neudecker, G.~P.~A. Nobre, G.~Noguere, M.~Paris, M.~T. Pigni,
  A.~J. Plompen, B.~Pritychenko, V.~G. Pronyaev, D.~Roubtsov, D.~Rochman,
  P.~Romano, P.~Schillebeeckx, S.~Simakov, M.~Sin, I.~Sirakov, B.~Sleaford,
  V.~Sobes, E.~S. Soukhovitskii, I.~Stetcu, P.~Talou, I.~Thompson, S.~van~der
  Marck, L.~Welser-Sherrill, D.~Wiarda, M.~White, J.~L. Wormald, R.~Q. Wright,
  M.~Zerkle, G.~\v{Z}erovnik, and Y.~Zhu.
\newblock {\em Nucl. Data Sheets}, 148:1 -- 142, 2018.

\bibitem{endf8_pfns}
D.~Neudecker, P.~Talou, T.~Kawano, A.C. Kahler, M.C. White, T.N. Taddeucci,
  R.C. Haight, B.~Kiedrowski, J.M. O'Donnell, J.A. Gomez, K.J. Kelly,
  M.~Devlin, and M.E. Rising.
\newblock {\em Nucl. Data Sheets}, 148:293 -- 311, 2018.

\bibitem{JEFF33}
OECD/NEA.
\newblock {JEFF}3.3 {E}valuated data {L}ibrary - {N}eutron {D}ata.
\newblock Technical report.

\bibitem{Coppola1970}
M.~Coppola and H.~H. Knitter.
\newblock {\em Zeitschrift f{\"u}r Physik}, 232:286--302, 1970.

\bibitem{Lovchikova1996}
G.~N. Lovchikova and A.~M. Trufanov.
\newblock In {\em Vop. At. Nauki Tekhn. Ser. Yadernye Konstanty}, page 102,
  1996.

\end{thebibliography}

\end{document}